\documentclass[manuscript,nonacm]{acmart}
\newcommand{{\proj}}[0]{PrivWeb}
\usepackage{graphicx}
\usepackage{tikz}
\usepackage{tabularx}
\usepackage{multirow}
\usepackage{multicol}
\usepackage[xcdraw,table]{xcolor}
\usetikzlibrary{trees}
\usepackage{subfig}
\AtBeginDocument{%
  \providecommand\BibTeX{{%
    \normalfont B\kern-0.5em{\scshape i\kern-0.25em b}\kern-0.8em\TeX}}}

\setcopyright{acmcopyright}
\copyrightyear{2018}
\acmYear{2018}
\acmDOI{XXXXXXX.XXXXXXX}

\acmConference[Conference acronym 'XX]{Make sure to enter the correct
  conference title from your rights confirmation emai}{June 03--05,
  2018}{Woodstock, NY}
\acmPrice{15.00}
\acmISBN{978-1-4503-XXXX-X/18/06}

\begin{document}

\title[Unobtrusive and Content-aware Privacy Protection For Web Agents]{PrivWeb: Unobtrusive and Content-aware Privacy Protection For Web Agents}

\author{Shuning Zhang}
\orcid{0000-0002-4145-117X}
\affiliation{%
  \institution{Tsinghua University}
  \city{Beijing}
  \country{China}
}
\email{zsn23@mails.tsinghua.edu.cn}

\author{Yutong Jiang}
\affiliation{
  \institution{Tongji University}
  \city{Shanghai}
  \country{China}
}
\email{jiangyytong@outlook.com}

\author{Rongjun Ma}
\affiliation{
  \institution{Aalto University}
  \city{Espoo}
  \country{Finland}
}
\email{rongjun.ma@aalto.fi}

\author{Yuting Yang}
\affiliation{
  \institution{Independent Researcher}
  \city{Ann Arbor}
  \country{United States}
}
\email{yutingyang986@gmail.com}

\author{Mingyao Xu}
\affiliation{
  \institution{University of Washington}
  \city{Seattle}
  \country{United States}
}
\email{mx37@uw.edu}

\author{Zhixin Huang}
\affiliation{
  \institution{Shantou University}
  \city{Shantou}
  \country{China}
}
\email{23zxhuang@stu.edu.cn}

\author{Xin Yi}
\orcid{0000-0001-8041-7962}
\authornote{Corresponding author.}
\affiliation{
    \institution{Tsinghua University}
    \city{Beijing}
    \country{China}
}
\email{yixin@tsinghua.edu.cn}

\author{Hewu Li}
\orcid{0000-0002-6331-6542}
\affiliation{
    \institution{Tsinghua University}
    \city{Beijing}
    \country{China}
}
\email{lihewu@cernet.edu.cn}

\renewcommand{\shortauthors}{Zhang et al.}


\begin{abstract}
While web agents gained popularity by automating web interactions, their requirement for interface access introduces significant privacy risks that are understudied, particularly from users' perspective. Through a formative study (N=15), we found users frequently misunderstand agents' data practices, and desired unobtrusive, transparent data management. To achieve this, we designed and implemented PrivWeb, a trusted add-on on web agents that utilizes a localized LLM to anonymize private information on interfaces according to user preferences. It features privacy categorization schema and adaptive notifications that selectively pauses tasks for user control over information collection for highly sensitive information, while offering non-disruptive options for less sensitive information, minimizing human oversight. The user study (N=14) across travel, information retrieval, shopping, and entertainment tasks compared PrivWeb with baselines without notification and without control for private information access, where PrivWeb reduced perceived privacy risks with no associated increase in cognitive effort, and resulted in higher overall satisfaction. 
\end{abstract}



\begin{teaserfigure}
    \centering
    \includegraphics[width=0.8\textwidth]{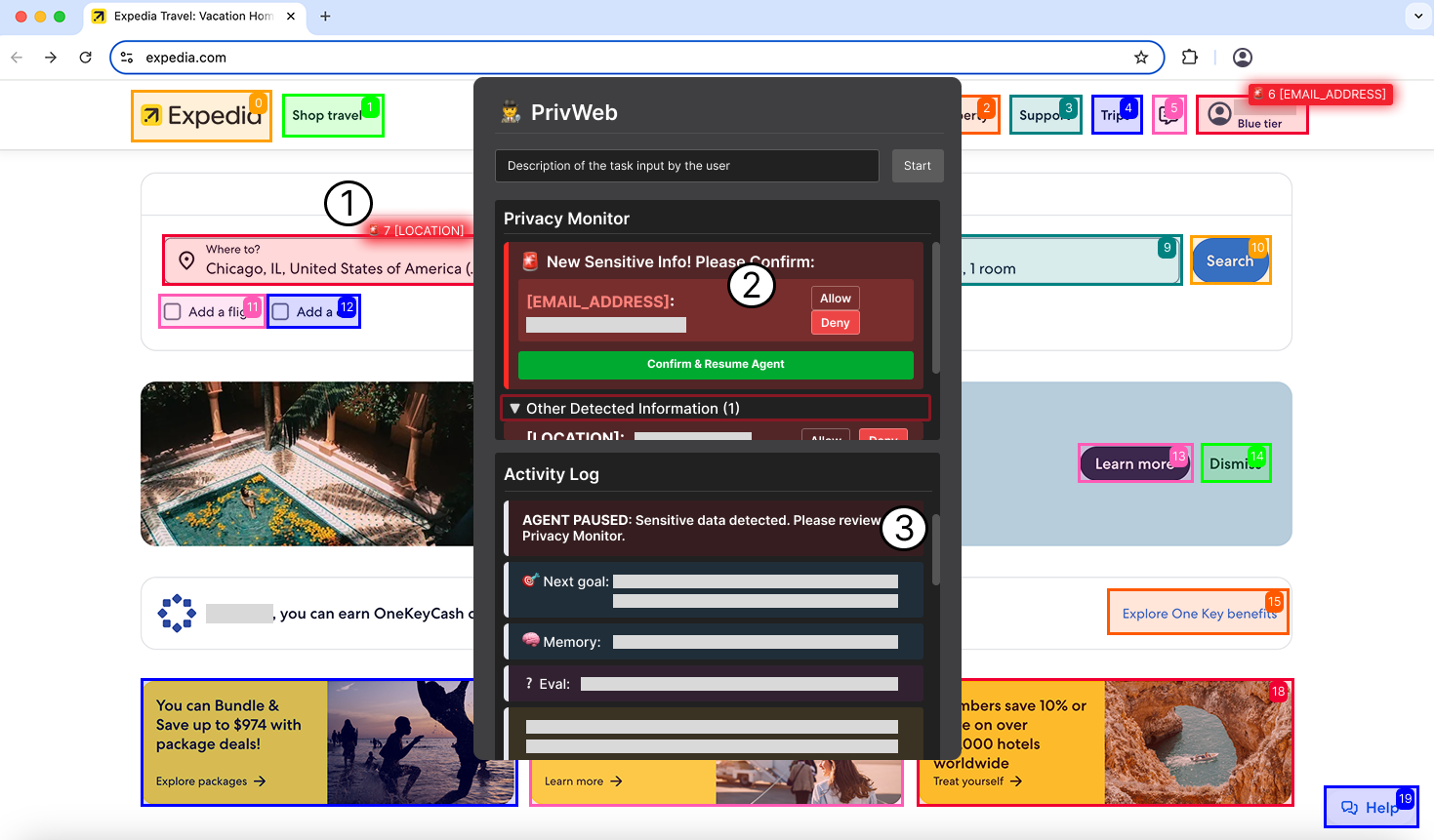}
    \caption{An illustration of \proj{}'s interface.  As a trusted, localized add-on for web agents, \textcircled{1} \proj{} uses in-situ highlighting to flag detected sensitive private information directly on the webpage. \textcircled{2} For private information with relatively low sensitivity, the panel shows the information and the control option, and adopts a privacy-by-default setting. For highly sensitive information, the agent's execution is paused, and a salient modal prompt requires the user to explicitly ``Allow'' or ``Deny'' web agents to use the detected information. \textcircled{3} The activity Log provides transparency into the agent's privacy control state, clearly indicating that it is paused, pending the user's privacy decision.}
    \label{fig:teaser}
\end{teaserfigure}


\maketitle

\section{Introduction}


Automating tasks through computers has long been a central goal in human–computer interaction (HCI)~\cite{bruns2009web,zhang2024large}. With the recent advances in Large Language Models (LLMs), this vision has extended into the realm of graphical user interfaces (GUI)~\cite{zhang2024large}, enabling agents to interact with digital environments in ways that closely mimic human behavior. To bridge natural language understanding and software manipulation~\cite{tang2025survey,cheng2024seeclick}, GUI agents integrate LLMs with techniques such as screen parsing, element detection, and action planning to enable end-to-end execution over interface-level operations~\cite{cho2024caap,lin2025showui}. GUI agents can operate across platforms including desktop, mobile environments ,and web browsers. Among them, web agents have shown strong potential in navigation, form filling, data extraction, and multi-step task automation across dynamic and visually complex websites~\cite{deng2023mind2web,lai2024autowebglm}.

However, web agents face significant risks~\cite{zhang2025characterizing} and drawbacks~\cite{chen2025toward}. Unlike traditional automation systems, \textbf{web agents often rely on screen capture or Document Object Model (DOM) inspection to perceive and manipulate interface content~\cite{humphreys2022data,liu2018reinforcement}, which grants them extensive access to potentially sensitive on-screen information~\cite{nguyen2024gui}}. This operational mode increases the privacy risks, especially when screenshots include personal data, confidential content, or third-party information, which may lead to unintentional privacy leakage or data exposure to external services.~\cite{chent2025toward,chen2025obvious}. Yet, how users perceive and make sense of these emerging privacy concerns is still largely underexplored. These risks are further compounded by the fact that web agents are typically designed to operate autonomously, without direct human oversight. The unpredictable behaviors of web agents poses challenges to traditional permission paradigms that were developed for structured API-based data access where users grant or deny resource requests via install-time or runtime prompts ~\cite{felt2012android,felt2012ask}. The operation of web agents across such diverse and dynamic settings~\cite{shi2025towards} thus demands novel approaches to supervision and privacy control. Therefore, this paper aims to first understand users' concerns and then design and develop a data protection technique that focuses on the interface-induced privacy issues of web agents, through answering the following research questions:




\textbf{RQ1.} What are users' understanding towards GUI agent's data practice, their current practices and challenges?

\textbf{RQ2.} How could we design a technique that unobtrusively visualize web agent's data practice and enable users to control?

\textbf{RQ3.} How do privacy notifications and controls affect users' behaviors, perceptions of privacy, and overall experience when interacting with web agents?

Towards RQ1, we first conducted a formative interview (N=15) to investigate users' understanding, practices, challenges and expectations towards web agents' data practice. We found users often could not understand the complicated data collection of web agents. They are also frustrated by the opaqueness of web agents' data practice. They hoped for a transparent communication of the web agents' data practice, as well as having control options for sensitive data.

Towards RQ2, we designed \proj{}, an in-situ notification plugin of web agents' interface-related data practice. \proj{} also facilitate users' control of web agents' collected data. \proj{} selectively highlights and anonymizes the sensitive data on the interface where web agents took operations, hinting users about potential data leakage without inducing cognitive overload. It also balances users' task flow through hinting in the banner, with peripheral pop-ups. The back-end of \proj{} monitors GUI agents' execution flow, extracting data collection and surfaces it to the front-end. In a technical evaluation, \proj{} achieved an average detection accuracy of 75.4\%, and specific classes like address, geo-location, demographic and health information achieving 100.0\% accuracy, suggesting the feasibility of on-device detection.

Towards RQ3, we conducted a in-lab evaluation study with 14 users, where we compared \proj{} with two alternatives in order to evaluate the individual effectiveness of \proj{}'s components. The two alternatives comprised one without notification or control of private information, mimicking the commercial systems (e.g., UI-TARS\footnote{\url{https://github.com/bytedance/UI-TARS}})that only provided task context. Another alternative is one with notification of users' private information usage but without control. We found that \proj{} significantly outperformed other techniques in protecting user privacy, while actively fostering users’ sense of agency and privacy awareness without imposing cognitive burden. To sum up, this paper makes three main contributions:  


$\bullet$ A formative study (N=15) identifying users' incomplete mental models of web agents' interface-related data practices, which creates a core conflict between their desire for granular control and the need to minimize cognitive load.
 
$\bullet$ Design and implementation of PrivWeb, a privacy protection add-on that provides unobtrusive notifications and selective user control over an agent's interface-induced data access.

$\bullet$ Technical evaluations and a user evaluation (N=14) showing that \proj{} mitigates users' privacy concerns, enhance their sense of control while balancing cognitive load, outperforming both alternatives.


\section{Background and Related Work}

This section reviews literature to addressing privacy challenges in GUI agents~\cite{zhang2024large}, especially web agents~\cite{deng2023mind2web,lai2024autowebglm}. We first introduce the backgrounds on GUI agents~\cite{deng2023mind2web,lai2024autowebglm}, then survey a spectrum of privacy risks, from flawed agent reasoning to adversarial attacks~\cite{liaoeia,chen2025obvious}. We finally examine established paradigms for notification and control in AI agents. 


\subsection{GUI Agents and Web Agents} 

With the advancement of web technologies, online services have deeply integrated into daily life, yet many web tasks, such as booking travel, managing information, or shopping, remain repetitive and time-consuming~\cite{ning2025survey}. Web agents, powered by large language models, offer a promising solution by automating such tasks through browsers, thereby enhancing productivity and efficiency~\cite{deng2023mind2web,yang2024agentoccam,zhuang2025workforceagent}. Recent work, such as \textit{MIND2WEB}, highlights both the feasibility and the growing research interest in building generalist web agents for diverse online environments~\cite{deng2023mind2web,lee2025learning}. These agents typically operate by parsing HTML-based observations (DOM trees or accessibility nodes), interpreting page structure and semantics, and issuing structured actions (e.g., click, fill, type) linked to unique element IDs~\cite{zhuang2025workforceagent,humphreys2022data,liu2018reinforcement}. Control mechanisms vary by design. Some agents run fully autonomously, while others adopt semi-autonomous paradigms that incorporate user-specified goals, action verification steps, or even “hybrid modes” to safeguard privacy. For instance, AgentSymbiotic proposes a privacy-aware hybrid architecture, where sensitive steps are offloaded to a local small LLM to minimize data exposure~\cite{zhang2025symbiotic}.

At the same time, these agents' inherent requirement for interface access still have unsolved privacy risks~\cite{mudryi2025hidden,zhang2025symbiotic}, making user-facing protections especially urgent. Their decisions are often driven by opaque planning modules, which may misinterpret page content and trigger unintended actions without explicit consent, such as revealing sensitive information or modifying user data~\cite{erdogan2025plan}. These risks highlight the need for safeguards like scoped execution, real-time feedback, and context-aware control to ensure secure deployment in dynamic web environments.

\subsection{Privacy Risks of GUI Agents} 

The increasing autonomy of GUI agents has introduced significant privacy risks, undermining user trust~\cite{zhang2024large}. These risks can be categorized into flawed agent reasoning, adversarial attacks, and vulnerabilities inherent to the agent-user interface, with the last category being particularly critical for web agents.

Agent may possess inherent weaknesses in privacy reasoning. They often require access to sensitive user data to function~\cite{edemacu2025privacy}, yet even state-of-the-art models can fail to handle this information appropriately, leading to inadvertent disclosures that a human would avoid~\cite{mireshghallahcan,shao2024privacylens}. This fundamental challenge is compounded by diminished user control and a lack of effective privacy guardrails~\cite{chen2025toward}.

A second class of risks involves external adversarial attacks, where malicious actors exploit the agent's operational context. These include fine-print injection attacks, which embed malicious content into a GUI to alter agent behavior~\cite{chen2025obvious}, and environmental injection attacks (EIA), which manipulate the underlying structured files like HTML, often in ways that are invisible to the user~\cite{liaoeia}.

Distinct from these external threats are privacy leakages inherent to an agent's standard operations on an interface. This risk arises when an agent, in the course of performing its duties, accesses or exposes sensitive data without malicious external influence. For example, agents with visual perception may inadvertently capture sensitive information through continuous screen recording~\cite{nguyen2024gui}. Similarly, agents that process structured data like HTML/DOM can access sensitive information embedded in the code that is not visible on the rendered page, creating a perceptual asymmetry between the user and the agent~\cite{nguyen2024gui}. While mitigation strategies like human-in-the-loop oversight have been proposed~\cite{openai2025operator, zhangagent}, managing the risk of inherent information leakage during routine web operations remains a primary challenge, which our work aims to address.

\subsection{Notification and Control in AI Agents}

Effective human oversight is critical for ensuring the reliable and safe operation of autonomous agents~\cite{chen2025toward}. This oversight is primarily enacted through mechanisms of \textit{notification and control}, which allow users to monitor, intervene in, and guide an agent's actions. The principles underpinning these mechanisms are well-established within the domain of privacy, which has a long history of research into the design~\cite{schaub2015design}, effectiveness~\cite{gluck2016short}, and standardization~\cite{kelley2010standardizing} of privacy notices. Foundational concepts such as \textit{notice and consent}~\cite{barocas2009notice} have shaped the design of user-centric control systems, a paradigm our work adapts for web agents. 

However, the high degree of autonomy and the sheer volume of operations in modern AI agents introduce new challenges for implementing effective oversight~\cite{chen2025toward}. The stream of an agent's outputs can overwhelm users, impeding their ability to make timely decisions or debug errors~\cite{zhang2025characterizing}. Research focusing on these problems identified prerequisites for effective oversight, such as ensuring the user has sufficient causal power and epistemic access~\cite{sterz2024quest}, and establishing alignment between the human and agent across multiple dimensions like knowledge and autonomy~\cite{goyal2024designing}. 

In response, recent work has focused on developing human-in-the-loop frameworks to facilitate manageable and collaborative privacy control with agents. Early efforts in this space sought to embed privacy awareness directly into LLMs to enable automatic control~\cite{shao2024privacylens}. Subsequent research has produced specialized tools for managing privacy in specific contexts, such as techniques for balancing privacy and utility when providing input to chatbots~\cite{zhou2025rescriber}, AI copilots that automatically edit images to reduce risks before social media sharing~\cite{monteiro2025imago}, and encryption-based methods to protect data during tool use~\cite{zhang2024privacyasst}. However, these solutions are tailored to text and social media images, which did not focus on interface-related privacy for web agents.

Concurrently, another stream of research has concentrated generally on LLM agents' oversight and control. For instance, some frameworks enable users to directly supervise and correct agent outputs~\cite{mozannar2025magentic}, while others structure the interaction loop around models of human cognitive dependency~\cite{yin2025operation}. Within the specific domain of web automation, systems like COWPILOT allow users to interleave their own actions with the agent's, overriding suggestions or resuming control as needed~\cite{huq2025cowpilot}. Although these systems exemplify the ongoing effort to balance agent autonomy with human control, their primary focus is on task correctness and execution rather than the specific privacy risks inherent in an agent's web-based operations, a gap this paper aims to address.


\section{Formative Study: Understanding Users' Privacy Concerns}

To understand users' practices and expectations regarding notification and control, we first conducted a formative study, with the dual aim to inspire \proj{}'s design. This approach allows us to ground our solutions in empirical evidence, a common adoption in privacy design~\cite{lee2024priviaware}.


\subsection{Participants and Recruitment}
We recruited 15 Chinese participants (14 males, 1 female) through distributing recruiting posters on social media platforms, including WeChat and RedNote. When recruiting, we used a screening survey consisting of which GUI agents participants have used, and for how long have they used the GUI agents, to ensure the quality of participants. The bias in gender is common in CS-related products and developing process~\cite{imtiaz2019investigating}. All reported familiarity with one or more systems, such as AutoGLM, Manus, Operator, Claude Computer Use, UI-TARS, or self-developed tools. The majority of participants (N=12) are in the 26-35 age range, with two aged 36-45 and one aged 18-25. Regarding educational background, seven held a Master's degree, while four held a PhD and four held a Bachelor's degree. The detailed demographics is shown in Table~\ref{tab:participants} in the Appendix. All interviews were conducted in Chinese, and each participant was compensated 90 RMB for their time.

\subsection{Interview Design and Procedure}

The semi-structured interview is designed around the central research question, divided into four parts: (1) users' basic demographics, (2) users' mental model about GUI agents' data practice, (3) their practices, controls and operations, and (4) their challenges and expectations. 
For (1), we asked about participants' usage experience of GUI agents and the tasks they usually use them for. For (2), we explored participants' perception of data practices, including collection, transmission, sharing, storage, and communication, drawing on dimensions highlighted in previous research~\cite{solove2005taxonomy,ma2025privacy}. We also asked about their opinions on the privacy risks of GUI agents. For (3), we asked about participants' current control strategies, along with the reasons and feelings behind the control. For (4), we asked participants about their expectations for privacy notification and control. We designed the interview questions around notification aspects of timing, content, and interaction mechanism, informed by the ``notice and consent'' guidelines~\cite{nissenbaum2011contextual}, and notification design framework~\cite{schaub2015design}. 
At the end of the interview, we invited participants to raise any additional topics of interest and freely share their opinions. All interviews were audio-recorded and automatically transcribed.

\subsection{Analysis Methods}
We adoptd Braun and Clarke's thematic analysis~\cite{braun2006using}. Before coding, all coders familiarized themselves with the interview transcripts and corrected the automatic transcriptions. After that, one primary author coded four randomly selected interview transcripts and drafted an initial codebook. All five authors then discussed the codebook, resolved potential questions, and reached an initial agreement.
The remaining interviews were distributed and coded independently by the five authors, with the codebook iteratively refined through intermittent discussions to resolve disagreements. 
Throughout the process, codes were merged and new ones proposed as findings emerged, with coders cross-checking each other’s work. All authors then collaboratively discussed and refined the themes.
We did not calculate inter-rater reliability, as we followed an inductive approach that emphasizes the process of theme generation and the benefits of multiple perspectives, rather than numerical agreement~\cite{mcdonald2019reliability}.

\subsection{Results}\label{sec:results}
To provide an overview of participants’ usage contexts, they reported mainly using GUI agents for technical development (e.g., code review for P1, P4, rename files for P1), data processing (e.g., create reports for P1, P11, P12, market research for P4, P8, P10, P14), information management (e.g., search literature for P1-3, P5, P8, P14, simple information retrieval for P1, P2, P5, P6, P14), and digital communication (e.g., post social media comments for P3, P10, P16).

\subsubsection{Understanding}

We first present participants' mental models towards the GUI agents' data practice, then detail their perceived privacy risks characterized by AI-specific vulnerabilities. Finally, we detail users' distrust evolving from both external and internal contradictory evidence.

\textbf{\textit{Incomplete mental models of GUI agents' data practice.}} 
In general, users expressed uncertainty about data practices, as they had not examined privacy polices or agreements (P8). Many had no clear idea about data practices, such as data collection or sharing. As P2 noted, \textit{``I'm unclear about this. To my understanding, they haven't specifically disclosed what information is transmitted to third parties.''} Some participants mistakenly believed GUI agents could collect internal company files stored on their laptops or expose financial information used for payment. For example, P2 worried that  \textit{``tracking residence locations or consumption habits allows estimation of economic status''}. Others speculated that GUI agents used their data for training or improving their products (P11), or uploaded and stored data on cloud servers (P14), though they had no evidence for these assumptions.

\textbf{\textit{Data collection from the interface as the main risk.}} Participants commonly believed that GUI agent collected their interface information, which could be misused or exposed. This included the screenshots, and also the specific information contained in the screenshots. They worried that the interface information may be used for purposes outside services, which could contain ample information. For example, P1 said, \textit{``The image may be used a second time or multiple times, or in other ways.''} Specific to web agents, participants also worried about the potential for scraping data in users' browsers, as P13 said, \textit{``It obtains all personal information such as the name. It feels like it can crawl all information [from the browser].''} On the interface, many participants also worried about their accounts and passwords, which could easily be extracted during the login process (P14) or in the website's main page.


\textbf{\textit{Perceived risks characterized by AI-specific vulnerabilities.}} These vulnerabilities included direct data collection~\cite{zhang2024s} (P1, P3, P6, P7, P8, P9, P10, P11, P14, P15), inferential risks~\cite{staab2024beyond} (P1, P5, P6, P8, P14) and their further misuse.
Participants mentioned external attacks and problems they perceived, including the exposure of private information, secondary use, and errors caused by AI.
For example, participants believed that the GUI agents could directly collect users' information (P6) or gather their private information unintentionally (P3). 
Others pointed to inferential risks (P8), with P6 noting,
\textit{``If you search for something on Ctrip, isn't that the perfect opportunity for Ctrip to gather your personal information?''} 
Beyond intentional collection, participants worried that AI models might collect data without permission (P5, P8), over-collect information, hallucinate (P16), or perform imprecise operations that cause privacy harms (P4).
As P1 expressed uncertainty, asking: \textit{``Will this image or interface be reused, either multiple times or for other purposes?''} Participants also raised concerns about external exploitation, with P4 hypothesizing,
\textit{``If a malicious actor gains control of your browser, they could potentially open a webpage or send an email, automatically logging into your email account. This could then be transmitted externally, putting your entire email account and password at significant risk.''} 

Participants held the opinion that these data may be misused, such as shared with third-party companies without users' consent (P1, P14), used multiple times (P1), or even sold illegally (P10). For example, P10 guessed, \textit{``Who else might this data be shared with? It's unclear, there could be third parties involved. Such information could very well be sold.''} P8 shared this concern, \textit{``It's exhausting to repeatedly specify my needs, only to have the system ignore them and reuse my data without consent.''}

\textbf{\textit{Distrust evolving from both external and internal contradictory evidences.}} Participants expressed a general distrust of GUI agents, influenced by social media, their judgment about the companies behind them, and their own usage experience. 
Participants who had seen similar software fail to protect privacy (P1) tended to perceive GUI agents as particularly risky.
This distrust was further amplified by public events (P2) or negative cases shared by others. Company reputation also played a role. As P2 noted,
\textit{``what people may place greater emphasis on is likely the company's reputation built over time''}.
Besides, participants emphasized the need for transparency; as P8 noted, they would  
\textit{``require transparent explanations for endorsement''}, and P2 adding they would \textit{``require the platform to provide technical explanations of their methods''}. 
Participants also evaluated companies’ technical capabilities (P2) and speculated about their intentions, with some suspecting that the company is deceiving them (P8).
During their practice, participants would judge based on their own usage scenarios (P8), 
noticed inconsistencies between agents’ behaviors and claims(P10), 
or observed other subtle issues that eroded trust.
Several participants also mentioned that trust takes time to develop
Collectively, these experiences suggest that trust in GUI agents is not only slow to build, but also fragile and easily undermined.


\subsubsection{Practice}

Participants' current practices could be divided into three interconnected parts: oversights, notification, and control, where participants dynamically oversee the execution process of agents. The agents notify users and users have control on specific data.

\textbf{\textit{Participants desire important control but skip unnecessary ones.}} Participants expressed that they desired the control over model actions and privacy protection. Some participants wanted the GUI agents to present screenshot content (P1), or want to control the models' operations and private information (P3). However, participants desired to bypass the login process, did not have the login process monitored (P1), and only desired formalized understanding (P2), but not real step-by-step guidance (P2). Overall, participants consistently emphasized the need for control over critical and sensitive aspects, while showing little interest in micromanaging routine or less important operations. 
Among all privacy-related elements, screenshots emerged as the most sensitive and frequently emphasized control point. For example, P1 argued that \textit{``The image is send back and forth for display, for example showing what was captured and whether it underwent any processing like blurring or pixelation to ensure the peace of mind.''} P2 also shared the opposite thoughts, clearly articulating this balance, \textit{``You might let him do certain things, but you wouldn't want to micromanage every single step.''}


\textbf{\textit{Current notify lack transparency and choice.}} Participants expressed the general distrust about the current notification (P4), stating that GUI agents did not present clear and publicized notification about their practice (P2). Participants lacked the knowledge about the privacy policy in general (P7), and also further lacked the choice towards the GUI agents' operations (P3). Specifically, some participants noted that privacy-related documentation exists but is inaccessible or impractical for ordinary users to read or evaluate. For example, P7 shared a common thought, \textit{``They might have documentation, but as a user, I'm never going to read it. I simply don't have the time, energy, or the expertise to judge how they're going to use my information—at least, they haven't made it clear.''} P3 empathized the lack of agency during usage, \textit{``It's not as if the user experience suggests that users have no choice and are just passively accepting it.''}


\textbf{\textit{Primary controls involved access control and information processing, though many participants expressed losing control.}} Participants commonly controlled the access of the GUI agents by enabling incognito browsing (P1), using separate VPNs and networks (P1), using sandbox-based installation (P4), and selectively set whitelist for different applications (P1). Participants also have specific information processing methods, including localized anonymization (P11), crafting fake data, and deletion. Some participants, specifically, expressed a lack of control, primarily originated from the in-transparency of the GUI agents (P1). For example, P1 explained that their main strategy was to remove residual data traces to prevent unintended access, \textit{``clear everything else, including browsing history, session data, and all traces to ensure that if the system tries to read my information, it can't retrieve much.''} P11 mentioned  anonymization in their workflow, \textit{``When writing work documents, personal names and organization names are coded, then replaced after writing.''} 


\subsubsection{Challenge and Expectations}

We presented the results around participants' expectation about desired control and notification, echoing the ``notice and consent'' framework~\cite{barocas2009notice,nissenbaum2011contextual}.

\textbf{Participants desired oversights on critical elements.} When asking participants about the timing about the oversight, they highlighted critical time point, including the monitoring the time when taking screenshots (P1), monitoring the content it copied, or payment related operations (P10). For example, P1 articulated the importance of active monitoring, ``\textit{All screenshot activities during the usage phase are actively monitored to track their precise interface origins.}''


\textbf{Participants expressed desire to control permission and context.} Participants desired to control contextual data (P3), and the permissions (P4). They also wanted to create separate accounts for control (P14). For example, P14 said, \textit{``I typically use it on my home PC, where I enforce physical environment isolation.''} 


\textbf{Participants preferred clear and to-the-point notification.} We separated the key of notification to three aspects. For notification manner, participants needed a clear notification in the privacy policy (P13), or marked out separately (P8). For notification timing, participants desired notifying before or after the usage (P9), but they did not desire that the notification interrupt their usage (P1). Some participants thought the notification should present both before and after the usage. Those who select presentation before the usage mostly cited reasons that they don't want to be disturbed during the process (P14). Participants also cited that if showed after the demonstration, they would feel of no use because of the demonstration (P1). They also cited that because before the execution, a proper notification could detail the process (P6). Those who select presentation after the usage mostly cited that the notification could act as a record or compliance (P2). The notification could also facilitate deletion afterwards (P3). Besides, participants also cited that although the notification should be presented afterwards, they are likely to omit the notification presented (P2). P13 emphasized the need for explicit alerts, \textit{``When a user enters a sensitive context, the system must display an alert, and require additional user interaction fro confirmation''} P8 echoed the importance of labeling, \textit{``For any data being collected, I believe it should be explicitly labeled/marked.''} P9 also articulated the preference for both timing and choice, \textit{``You always seek my confirmation before usage, and afterwards I retain the ability to revise my preferences.''} P14 argued for efficiency considerations, \textit{``Operational efficiency is critical. Real-time notifications during the process would likely degrade system performance.''} 

\textbf{Participants desired information around privacy and contexts.} For privacy related information, participants mostly desired the description around data usage, which portion of the data users could opt-in or opt-out, the data collected by the model, and what specification the model has done to the data, like anonymization or redaction (P3). They are also curious about the model's capacity border, and the summarization of the current action of the model (P2). For example, P3 highlighted the need for explicit disclosure, \textit{``Upon task completion, the system should explicitly disclose the data categories collected and required permissions with contextual justification.''} 


\section{Design and Implementation of \proj{}}\label{sec:design_implementation}

This section introduces the threat model, design rationale and implementation details for \proj{}, a technique acting as a system service, which enhance users' awareness and control of web agents' interface-related privacy disclosures.

\subsection{Threat Model}

We consider a threat model centered on the interaction between a user and an LLM-powered web agent operating within their web browser. We assume the user's client machine and web browser are secure and not compromised. The web agent is considered ``honest-but-curious'': its primary objective is to complete the user-assigned task by interacting with web interfaces, for which it has privileged access to content rendered on the screen and the underlying DOM. The central privacy threat originates from the agent's potential to over-collect, mishandle, or inadvertently leak sensitive user information, such as login credentials, personal identifiers, and financial details, to third-party services or its own developers during its operation. This model is common both in the current privacy-related research in GUI agents~\cite{zharmagambetov2025agentdam} and also for traditional app developers~\cite{venugopalan2024aragorn}. Our model, therefore, focuses on protecting the user from the agent itself, which may lack the necessary privacy awareness or make unsafe inferences. Threats from a compromised operating system, malicious browser extensions, or network-level attacks are considered out of scope.

\subsection{Design Goals}

The primary goal of \proj{} is to empower users to comprehend and manage their interface-related privacy exposure during the interaction with web agents. To this end, we formulate the following design goals:

\textbf{DG1. Foster comprehensive situational awareness of privacy state.} Effective privacy management requires not just access to information, but a clear and ongoing understanding of one's current status, a concept known as situational awareness. This goal focuses on translating abstract data permissions into a tangible, readily understood representation of the user's privacy posture. The design should provide an ambient and intuitive understanding of ``where the user is'' and ``how they got there'' in terms of their data disclosure trajectory, thereby enabling informed forward-planning within an interactive task.

\textbf{DG2. Empower granular and effective control.} We followed the ``notice and consent'' mechanism~\cite{nissenbaum2011contextual}, which were echoed in the formative study that participants highlighted the control over the interface-related privacy. 

\textbf{DG3. Balance control with cognitive load.} Our study revealed a significant paradox: while users express a desire for fine-grained control over their personal data, they also fear to be overwhelmed by the cognitive demands of making frequent, granular privacy decisions. This phenomenon, echoing ``privacy fatigue''~\cite{choi2018role}, leads to users defaulting to permissive behaviors simply to reduce mental effort. Consequently, a primary goal of \proj{} is to provide meaningful control at critical design points, while abstracting less consequential choices to mitigate user fatigue, therefore empowering users without overburdening them.

\subsection{Interface Design}

The \proj{} interface is designed to provide users with continuous situational awareness of their privacy state without disrupting their workflow. This is achieved through a multi-component, in-situ display system (see Figure~\ref{fig:teaser}). The main interface is a movable side panel that, by default, mimics a working panel, allowing for ambient monitoring. 

The panel itself is structured with a task context panel and a privacy panel (see Figure~\ref{fig:teaser} \textcircled{2}). The privacy panel is organized hierarchically to reflect a data sensitivity schema derived from our formative study. The upper portion, the most prominent area, is reserved for highly sensitive information, such as financial and account details, to ensure immediate user attention. Below this, a lower section displays information of lower sensitivity, such as names and addresses. This design balances the need for privacy with the demand for utility, preventing constant interruptions for less critical data points. 

Beyond the panel, \proj{} provides in-situ highlighting directly on the webpage (see Figure~\ref{fig:teaser} \textcircled{1}). When sensitive data is detected, \proj{} highlights the element with a bounding box and an additional marker. This visual cue remains for about three seconds, providing a brief but clear indication of the specific data \proj{} has identified and is currently anonymizing. This approach provides a tangible, real-time representation of the user's data trajectory, addressing the need for transparent communication identified in our formative study.

\subsection{Interaction Design}\label{sec:interaction_design}

\proj{}'s interaction design is fundamentally rooted in a privacy-by-default design~\cite{cavoukian2009privacy}, which mandates that all sensitive information is blocked from agent access unless the user explicitly grants permission . This approach directly addresses the issue of ``privacy fatigue''~\cite{choi2018role} by shifting the burden of control away from the user and minimizing cognitive load, a key tension identified in the formative study . For privacy controls, \proj{} provides a binary choice for users to either ``Allow'' or ``Deny'' the collection of information. This simplified control mechanism was adopted because users often find it difficult to understand and implement more complex, fine-grained controls, and it avoids the need for users to grasp intricate data obfuscation techniques, like those in a differential privacy framework~\cite{nanayakkara2023chances,cummings2021need}.

\proj{}'s interaction logic is driven by a two-tiered classification scheme for sensitive information. Rescriber~\cite{zhou2025rescriber} categorized several classes of Personal Identifiable Information (PII) for the input to LLMs: \textit{name}, \textit{address}, \textit{email}, \textit{phone number}, \textit{ID}, \textit{online identity}, \textit{geo-location}, \textit{affiliation}, \textit{demographic attribute}, \textit{time}, \textit{health information}, \textit{financial information}, \textit{educational record}. It encompassed the recent work on GUI agents~\cite{zharmagambetov2025agentdam}, and represented a finer classification than AgentDAM. Notably, AgentDAM mentions six classes of sensitive information: \textit{personal and contact information}, \textit{religious, cultural, political ... identification}, \textit{employer and employment data}, \textit{financial information}, \textit{educational history}, and \textit{medical data}. From the formative study, we also did not identify that participants uttered other information types specific to screenshot data on their accomplished tasks in web browsing except the above, therefore we decided on the classification of PIIs as shown in Table~\ref{tab:user_preferences}. According to the formative study, we set on those four classes as needing explicit control, which means intrusive reminding, \textit{financial information}, \textit{ID}, \textit{online identity} and \textit{email}. For other information, they only triggered unobtrusive notification, but not interruptive notification.

\begin{table}[h!]
\centering
\caption{The classification of PIIs for different privacy control and notification settings.}
\label{tab:user_preferences}
\begin{tabular}{p{0.4\textwidth}|p{0.25\textwidth}|p{0.25\textwidth}}
\toprule
\textbf{High sensitive (Explicit control)} & \textbf{Medium sensitive} & \textbf{Low sensitive} \\
financial information & name &  geo-location\\
ID & address & affiliation \\
online identity & phone number & time \\
email & demographic attribute & educational record\\
& health information & \\
\bottomrule
\end{tabular}
\end{table}

Beyond this two-tiered classification schema, \proj{} employs a three-level sensitivity ranking to further assist user decision-making (see Table~\ref{tab:user_preferences}), a classification inspired directly by our formative study. This ranking classifies information as either highly, medium or low sensitive. \textit{Financial information}, \textit{ID}, \textit{online identity}, and \textit{email} are ranked highly sensitive. \textit{Name}, \textit{address}, \textit{phone number}, \textit{demographic attribute}, and \textit{health information} are ranked as medium sensitive. \textit{Affiliation}, \textit{educational record}, \textit{time} and \textit{geo-location} are ranked as low sensitive. This classification does not imply that low-sensitivity information is not sensitive, but rather prioritizes user attention toward the most critical data points. All detected information is then presented to the user in the privacy panel, ranked by its sensitivity from top to bottom. To provide a clear visual cue, we used a color-coding system, where low-sensitivity information is highlighted in yellow, medium-sensitivity in orange, and high-sensitivity information in red, effectively warning users and guiding their focus. 

To enhance transparency and user control, the interface also displays every element the agent interacts with, providing a clear view of the agent's task progress. Users can directly act on any displayed element. Clicking an element triggers its immediate and persistent anonymization, preventing its contents from being transmitted to the web agent in any subsequent interactions. This manual override functions as a final safeguard, empowering users to correct for any automated mis-detections and exercise final authority over their data disclosure.


\subsection{Algorithm Design}

\proj{} acts as a privacy-preserving intermediary between a user's browser and the web agent. Its core function is to intercept data requests from the agent and programmatically filter out sensitive information before transmitting it to the web agent. This process is triggered whenever a web agent requests interface information, such as the DOM tree or a screenshot.

\begin{figure}[!htbp]
    \includegraphics[width=\textwidth]{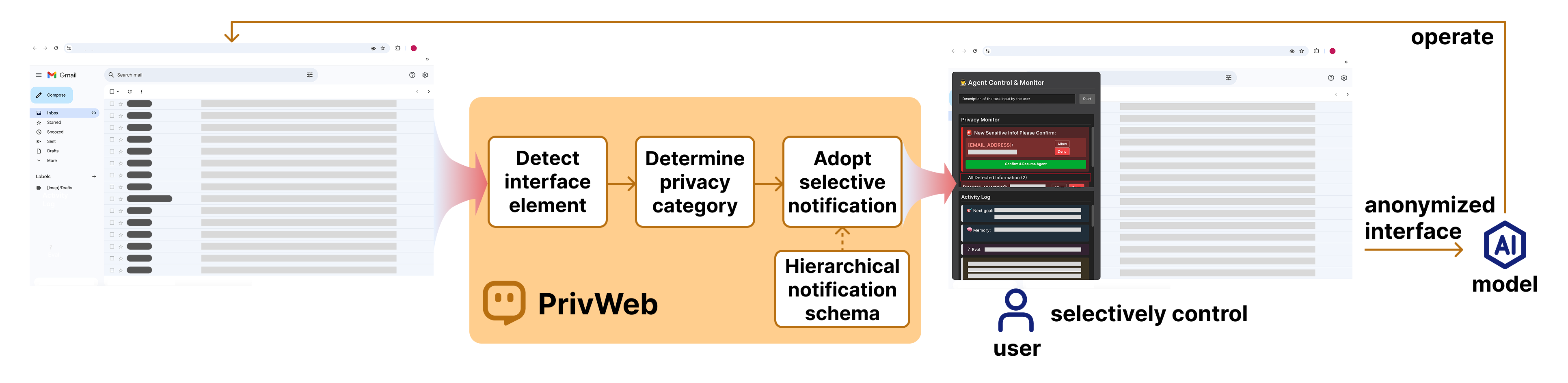}
    \caption{The PrivWeb architecture, which protects user privacy by detecting sensitive interface elements, engaging the user with selective, hierarchical notifications for granular control, and subsequently providing an anonymized interface to the web agent model for task execution.}
    \label{fig:algorithm}
\end{figure}

The algorithm follows a multi-step process for privacy protection. First, \proj{} uses a browser automation framework, such as Playwright, to acquire the DOM tree and parse all text-containing elements such as \verb|<div>|, \verb|<span>|, \verb|<a>|, \verb|<input>|, \verb|<textarea>|, and \verb|<select>|. It systematically extracts all associated sub-elements with texts, including visible text, accessibility labels, placeholders, and values, and assigns a unique ID to each element-information pair for subsequent redaction. 

Next, \proj{} sends these extracted elements to a localized LLM, specifically Qwen3-8b, to recognize and classify private information. A local LLM was chosen to ensure the system remains a trusted, on-device service, thereby avoiding data transmission to cloud-based models. The LLM outputs a structured list of detected private information, including its privacy class and the unique ID that links it back to the original interface element.

Upon receiving the LLM's output, \proj{} utilizes the unique IDs to locate the original elements and redact them. Redaction is achieved in two ways: either by deleting the web element from the DOM tree before it's sent to the agent or by deleting the element and re-rendering the interface to create a screenshot that is then sent to the agent. Simultaneously, to provide real-time user feedback and enhance transparency, \proj{} highlights the detected sensitive element on the user's screen for about three seconds. This is accomplished by injecting a bounding box into the DOM tree that precisely encloses the element. For highly sensitive data, an additional marker is added to the bounding box to explicitly remind the user of its presence. The system's sensitivity mapping and the visual demonstration on \proj{}'s panel are automatically performed via a pre-defined schema, based on the sensitive information's class outputted by the LLM. 

\subsection{Implementation Details}\label{sec:implementation_details}

We implemented PrivWeb as a system service using a series of integrated components. We deployed the Qwen3-8b model for private information extraction, utilizing Ollama for local execution with its default parameter settings to ensure on-device privacy. The prompt for privacy detection included (1) the task's aim, which is to detect PIIs, and output the PII's class in conjunct with the PII, (2) definition of PII, (3) the descriptions of the private information categories, and (4) rules that LLMs needed to obey for detection, such as what information to exclude. We provided the detailed prompt for the implementation \textit{in the supplementary materials}. Communication between the service and the user interface is handled via socket connections. We used Pywebview for the GUI demonstrating the privacy controls. Additionally, we used system APIs for capturing screenshots of the active browser window. We used Playwright for web automation and programmatic DOM tree manipulation. 

\subsection{Technical Evaluation}

We conducted a technical evaluation to examine the privacy protection effectiveness of the anonymization. Following prior work~\cite{chen2025obvious}, we curated a dataset by referring to the Mind2Web benchmark~\cite{deng2023mind2web} and crafted 100 tasks, ensuring coverage across diverse context such as entertainment, work, travel, shopping and information retrieval. To generate realistic data, experimenters executes these tasks using web agents on their own laptops, providing authentic information like account credentials as needed. Experimenters used different accounts to diversify the dataset. This effort resulted in 107 sessions\footnote{Note that for a few tasks we recorded more than one session.} and 2,041 steps, averaging 19.07 steps per task (SD=14.59). Through human annotation, we identified 2,189 unique instances of sensitive information entity following the classification outlined in Section~\ref{sec:interaction_design}. The uniqueness referred to that if exactly the same information appears in multiple steps of a task, we excluded that information. These information spanned all our defined categories, with the least category having at least ten instances, ensuring a robust evaluation. The complete distribution of these private information is detailed in Appendix~\ref{app:distribution}. We evaluated the models introduced in Section~\ref{sec:implementation_details}, alongside other models, to compare the accuracy and latency across different architectures and interface information types (see Tables~\ref{tab:accuracy_information} and ~\ref{tab:latency} for results).

\begin{table}[!htbp]
\centering
\caption{Accuracy for different information types for different models.}
\label{tab:accuracy_information}
\begin{tabular}{@{}llll@{}}
\toprule
 & gpt-4o & qwen3-8b & qwen3-4b \\ \midrule
 Name & 77.8\% & 75.0\% & 70.0\% \\
 Address & 100.0\% & 100.0\% & 75.0\% \\ 
 Email & 83.3\% & 85.7\% & 78.6\% \\
 Phone number & 100.0\% & 92.9\% & 87.5\% \\ 
 Online identity & 76.3\% & 71.1\% & 52.1\% \\ 
 Geo-location & 100.0\% & 100.0\% & 61.8\% \\ 
 Affiliation & 76.9\% & 63.0\% & 52.4\% \\ 
 Demographic attribute & 100.0\% & 100.0\% & 80.0\% \\ 
 Time & 100.0\% & 83.3\% & 75.0\% \\ 
 Health information & 100.0\% & 100.0\% & 83.3\% \\
 Financial information & 100.0\% & 100.0\% & 75.3\% \\ 
 Educational record & 74.3\% & 70.7\% & 54.6\% \\
 ID & 82.2\% & 72.5\% & 41.7\% \\
 Avg. & 82.1\% & 75.4\% & 58.9\% \\
 \bottomrule
\end{tabular}
\end{table}

\begin{table}[!htbp]
\centering
\caption{Average latency (in seconds) for detecting all information per DOM page across different models.}
\label{tab:latency}
\begin{tabular}{@{}lllll@{}}
\toprule
 & deepseek-7b & qwen3-8b & qwen3-4b & qwen3-1.7b \\ \midrule
 Latency & 8.52 & 6.42 & 4.05 & 2.86 \\
 \bottomrule
\end{tabular}
\end{table}

These results suggested that the anonymization models are capable of protecting users' privacy on interfaces. As shown in Table~\ref{tab:accuracy_information}, qwen3-8b model achieved a robust average accuracy of 75.4\%, while gpt-4o performed even better at 82.1\%. The models exhibited exceptional accuracy for several highly sensitive information types, reaching 100\% for categories such as \textit{address}, \textit{geo-location}, \textit{demographic attributes}, and \textit{health information}. While accuracy was comparatively lower for more nuanced types like \textit{online identity} and \textit{educational records}, the qwen3-8b model still maintained a respectable performance of over 70\% in these areas. In contrast, the smaller qwen3-4b model's average accuracy of 58.9\% indicates a significant performance degradation that may render it unsuitable for reliable deployment.

Our latency analysis (Table~\ref{tab:latency}) found the average latency per DOM page ranged from 8.52 seconds for the larger deepseek-7b model to just 2.86 seconds for the most qwen3-1.7b model, with qwen3-8b showing a latency of 6.42s for processing the DOM page, which is acceptable and did not affect user experience in the subsequent user study.


\section{User Study: Evaluating \proj{}'s Effectiveness in Empowering Users' Privacy Management}

We conducted a user study to examine whether \proj{} could (1) help users complete the task and manage the privacy effectively, efficiently and with low cognitive load. We further (2) examined users' behavior during the usage of web agents, especially in the aspects of privacy.

\subsection{Participants and Apparatus}

We recruited 14 participants (8 males, 6 females, with a mean of 23.2, SD=2.6) through distributing the posters online in WeChat groups. We did not require participants to have prior experience with web agents. Of them, 2 participants are from computer science related occupations, 3 participants are from engineer occupations and other participants are from multiple disciplinary. 4 participants were with a master's degree or higher, with 7 participants holding a bachelor's degree, and 3 participants with a high school degree. The study was approved by our Institutional Review Board (IRB), and each participant was compensated 200 RMB. The experiment consisted of two parts, for which participants used either their personal devices or a provided Lenovo R9000P laptop, to reduce the effect of device settings.

\subsection{Study Design}

The study used a within-subjects design, with technique as the only factor. We compared our technique with other alternative techniques, which removed the notification module, or removed the control module, or removed both. We detailed the settings below: 

$\bullet$ \proj{}: implemented with control and notification function, as in Section~\ref{sec:design_implementation}. 

$\bullet$ Without control (denoted n/c): we removed the control options, where participants could only see the notification but could not control the agents. 

$\bullet$ Without both (denoted n/b): we removed both the control and notification options, where participants could not see the private information highlighted on the interface. However, in all the above settings, the participants could see the status of the task. This mimicked the current commercial products like UI-TARS~\footnote{https://github.com/bytedance/UI-TARS}.

We selected the tasks from Mind2Web dataset~\cite{deng2023mind2web} to ensure coverage of diverse web-based activities. For each task, participants were given a background scenario and a pre-written prompt designed to guide th web agent, such as ``Access institutional finance portal to download payment receipts for student fees''. Participants were instructed that they were free to modify the provided prompt or compose their own to achieve the task objective. The complete list of tasks is available in Appendix~\ref{app:task_set}. To evaluate user experience, especially regarding privacy protection and control, we measured the following aspects, drawing from practices in prior work~\cite{ahmad2022tangible,zhou2025rescriber}:



$\bullet$ Cognitive load: we used NASA-TLX to measure the cognitive load developed by Hart and Staveland~\cite{hart1988development}. 

$\bullet$ Trust: Trust was assessed using the \textit{Trust in Automation Scale} proposed by Jian et al.~\cite{jian2000foundations}, which evaluates users’ confidence in the behavior and reliability of automated systems. 

$\bullet$ Perceived control: Perceived control was measured through items adapted from the PCTL scale, initially introduced by Xu~\cite{xu2011information} to capture users’ perceived agency over the collection and use of their personal data. 

$\bullet$ User experience: User experience was measured using the UMUX-Lite scale developed by Lewis and Sauro~\cite{lewis2013umux}, which provides a concise and reliable assessment of user experience.

$\bullet$ Perceived privacy protection: We adopted the entries from Zhou et al.~\cite{zhou2025rescriber}, measuring whether each technique could \textit{``reduce the disclosure of unnecessary information''} (\textit{unnecessary information protection}), \textit{``reduce the disclosure of personal information''} (\textit{personal information protection}), result in \textit{``fewer privacy concerns using the technique''} (\textit{reduce concern}) and whether users \textit{``would like to the technique''} (\textit{willingness to use}).

All scales were measured using standard 7-point Likert formats. We additionally conducted semi-structured interviews after the experiment, following prior practices~\cite{zhou2025rescriber,zhang2024privacy}. We started with introduction and overall impressions, allowing the participants to recall the most salient experience without immediately priming them on specific topics like ``privacy'' or ``control''~\cite{zhang2024privacy}. We then focused on different aspects, first asking whether users were aware of the privacy during the execution and their oversight. We also asked them about their control and agency. We further probed them about the cognitive load, trust and potential negative impacts, which were also important especially for the execution of agents~\cite{zhang2025characterizing}. We finally asked participants to compare different techniques, suggest improvements and have an open-ended sharing of their experience.


\subsection{Procedure}

The experiment was divided into two parts, with the order counterbalanced across participants~\cite{edwards1951balanced}. For the first part, participants mimicked their daily usage setting, and for the second part they used the laptops provided by the experimenter in a controlled setting. In each part, participants completed one task for each of the three techniques. These totaled six tasks for two parts. The specific task were randomly assigned from the predefined pool, and the presentation order of the techniques was randomized. For each task, participants initiated a web agent by entering a prompt.  We did not force participants to monitor agents' execution. They were allowed 15 minutes per task, and could attempt multiple times until they considered it complete. If the task was not completed within this period, they needed to shift to the next task, to avoid spending too much time on a single task which may not be completed by web agents. After each task, participants filled out a questionnaire. Upon completing all six tasks, they took part in a final exit interview. 

\subsection{Results}

For the analysis on subjective ratings, as these data did not conform to normal distribution, we adopted Friedman's non-parametric test for analysis, with Nemenyi test for post-hoc analysis. We used thematic analysis~\cite{braun2006using} on the transcribed data, one primary author inductively open-coded a subset of transcripts (3 transcripts) to develop a preliminary codebook. This initial codebook was then collaboratively discussed, reviews and refined by the primary author and a secondary author to resolve discrepancies, and establish clear definitions for each code. Subsequently, the primary author coded the remainder of the transcripts, with intermittent discussions with the secondary author to ensure quality. They subsequently collaboratively aggregated codes into themes. 

\subsubsection{Subjective Ratings}

\begin{figure}[!htbp]
    \centering
    \includegraphics[width=0.8\textwidth]{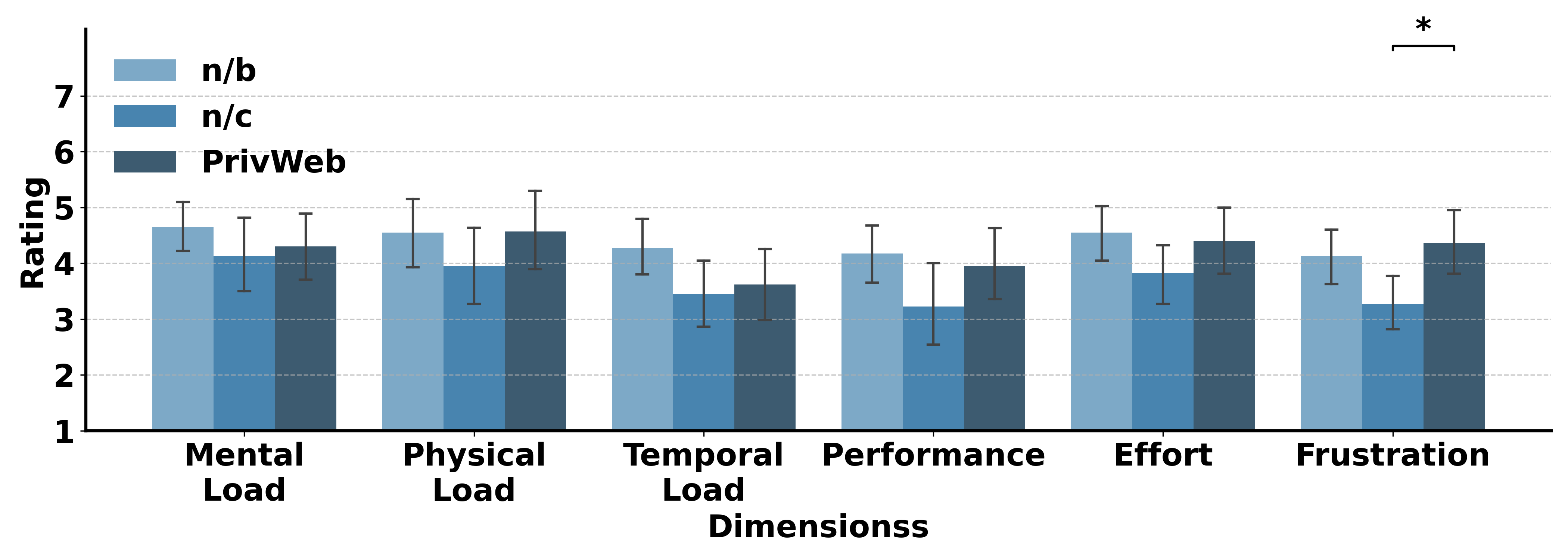}
    \caption{Participants' NASA TLX ratings across different techniques (1: most negative, 7: most positive). Errorbar indicated one standard deviation. * indicated significance at $p < .05$.}
    \label{fig:nasa_tlx}
\end{figure}

Analysis of the subjective ratings reveals a nuanced perspective on the user experience across the three techniques, as shown in Figure~\ref{fig:nasa_tlx}. Regarding cognitive load, as measured by the NASA-TLX scale, we found no statistically significant differences across five of the six dimensions: \textit{Mental Demand} ($\chi^2_2 = 1.273$, $p = .53$), \textit{Physical Demand} ($\chi^2_2 = 4.044$, $p = .13$), \textit{Temporal Demand} ($\chi^2_2 = 1.235$, $p = .53$), \textit{Performance} ($\chi^2_2 = 1.870$, $p = .39$) and \textit{Effort} ($\chi^2_2 = 0.429$, $p = .80$). A significant difference was found for \textit{Frustration} ($\chi^2_2 = 8.346$, $p = .015 < .05$, Kendall's $W = .298$), where post-hoc tests found significant difference between \textit{\proj{}} and \textit{n/c}.  This suggests \proj{}, including both notification and control mechanisms, do not impose a significant additional cognitive burden on the user compared to the alternative conditions. In contrast, when not granted with control but with privacy notifications, participants may have further anxiety and frustration.

Our study demonstrated significant effects in dimensions directly related to trust and privacy protection, suggesting the feasibility of \proj{}. We found a significant effect of the technique on user \textit{Trust} ($\chi^2_2=7.000$, $p=.030<.05$, Kendall's W = .250). More profoundly, we identified strong significant effects on all three privacy-centric dimensions: \textit{unnecessary information protection} ($\chi^2_2 = 11.375$, $p = .003 < .01$, Kendall's W = .406), \textit{personal information protection} ($\chi^2_2 = 14.816$, $p < .001$, Kendall's W = .529), and the ability to \textit{reduce concern} ($\chi^2_2 = 16.148$, $p < .001$, Kendall's W = .577). Post-hoc comparisons confirmed that \textit{\proj{}} was significantly preferred over the alternative \textit{n/b} condition across all three of these dimensions (\textit{unnecessary information protection}: $p < .01$, \textit{personal information protection}: $p < .01$, \textit{reduce concern}: $p < .001$). Furthermore, \textit{\proj{}} was also rated significantly higher than the notification-only \textit{n/c} condition for both \textit{unnecessary information protection} ($p < .01$) and \textit{personal information protection} ($p < .05$). This strongly suggests that the combination of transparent notification and actionable control is the critical factor in empowering users. Merely providing awareness of data collection without control (\textit{n/c}) is demonstrably less effective.

However, Our analysis revealed no statistically significant differences among the techniques for several user experience metrics. These included \textit{Perceived Control}, as measured by the PCTL scale ($\chi^2_2=5.778$, $p=.06$), overall \textit{User Experience}, measured by UMUX-Lite ($\chi^2_2 =4.500$, $p=.10$), self-reported \textit{Willingness to Use} ($\chi^2_2 = 1.469$, $p = .50$). This suggests that while \proj{} introduced new privacy control features, it did not negatively impact the general user experience or measurably alter users' sense of control within the scope of this study. The lack of significance in these areas may indicate that constructs like perceived control may be correlated with other factors, or could require longer-term interaction to be influenced.

\begin{figure}[!htbp]
    \centering 
    \includegraphics[width=0.8\textwidth]{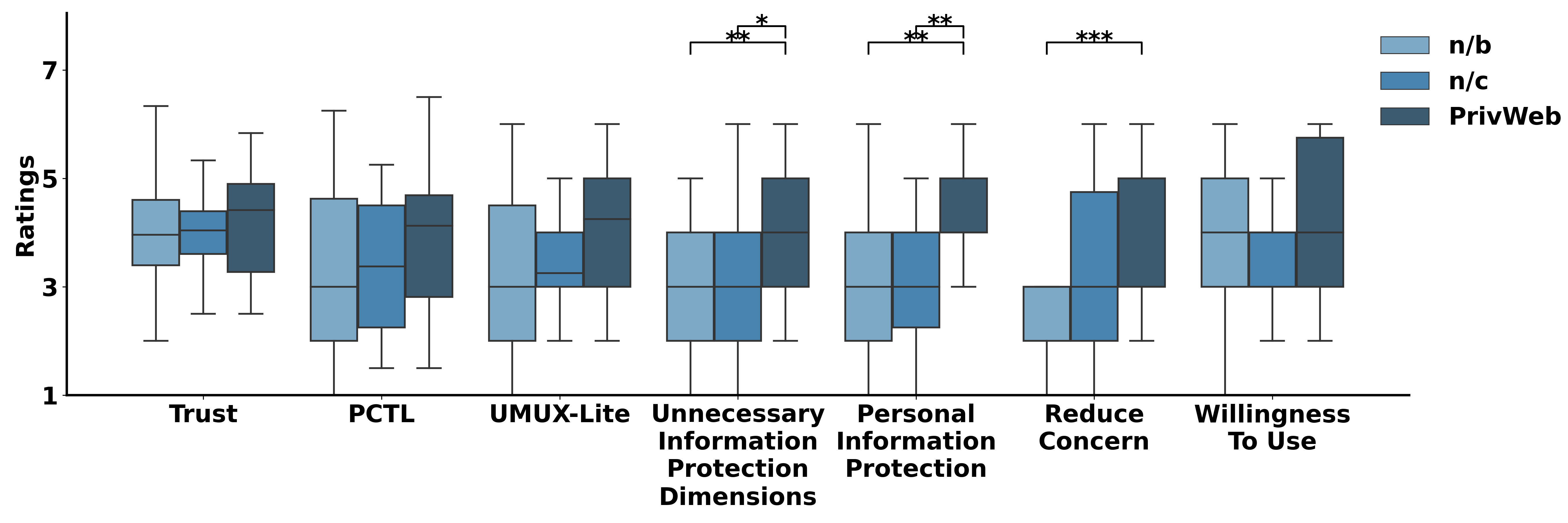}
    \caption{Participants' subjective ratings regarding the Trust, PCTL, UMUX-Lite, Unnecessary information protection, Personal information protection, Reduce concern and Willingness to use dimensions across different techniques (1: most negative, 7: most positive). *, **, *** separately indicated significance at $p < .05$, $p < .01$ and $p < .001$.}
    \label{fig:combined_scales}
\end{figure}



\subsubsection{Subjective Feedback}\label{sec:subjective_feedback}

Our analysis synthesizes user feedback in interacting with different techniques. Overall, an overwhelming majority of participants (13 out of 14) preferred \textit{PrivWeb} for their personal use. Specifically, we synthesized four themes that illustrating the nuances of privacy protection, control and trust: (1) the foundational role of user control and transparency in establishing trust, (2) the inherent tension between the desire for granular privacy management and the associated cognitive load, (3) the nuanced and context-dependent nature of the user's privacy calculus, and (4) the overriding impact of agent competence and predictability on the user-agent relationship.

\textbf{Importance of control and transparency in building trust.} The most critical factor in fostering user trust is the provision of explicit mechanisms for both awareness and oversight. Participants overwhelmingly preferred \textit{\proj{}} that combined transparency (notifications) with actionable controls (the ability to grant or deny access), perceiving this combination as a prerequisite for psychological safety. The sense of control was not a supplementary feature but a core component of trust, directly mitigating feelings of vulnerability.

\textit{Control as a foundation for safety:} The ability to actively manage the agent's access to information was directly equated with a sense of security. As one participant stated, \textit{``Having the control function definitely makes me trust it more... It gives me a sense of choice... This makes me feel safer''} (P13). This sentiment was echoed by another user who valued the ability to manage potential information leaks: \textit{``I feel that having a mechanism that allows me to control it gives me a greater sense of security. At least the information that might be leaked is controllable''} (P7).

\textit{The anxiety of awareness without agency:} The findings reveal a critical design principle: awareness without control is more detrimental to the user experience than a complete lack of awareness. The condition that provided notifications about privacy-implicating actions without offering a way to intervene induced significant anxiety and helplessness. One participant articulated this frustration vividly: \textit{``It only tells me that an operation involving privacy is about to be performed, but I have no way to stop it... it's quite uncomfortable. It's almost better not to tell me anything at all.''} (P7)

\textit{Transparency as a window into the ``black box'':} Users valued notifications and controls because they demystified the agent's operations. This transparency into the agent's decision-making process was crucial for building confidence. As P10 noted, controls \textit{``let me know that specific information it has read, rather than it being a black box where I only know the result but not what it did.''}

\textbf{Tension between granular control and cognitive load.} While users universally desired control, its implementation presented significant usability challenges, creating a trade-off between privacy protection and cognitive effort. 

\textit{The burden of constant decision-making:} An excessive number of interruptions for privacy decisions was perceived as a significant burden that undermined the agent's primary purpose of automation. This high cognitive load led to frustration and, in some cases, the abandonment of meticulous privacy management in favor of task completion. As P4 stated, \textit{``if I have to constantly supervise it during the process, it will consume some of my energy, and I might as well do it myself.''} Another participant worried that frequent prompts \textit{``might bring a certain burden.''} (P8)

\textit{The criticality of actionable, legible information:} The effectiveness of any privacy control system is contingent on the clarity of the information presented. Users often ignored or misunderstood notifications that were overly technical, such as raw code or URLs, rendering the controls ineffective. In contrast, clear, human-readable labels, such as ``Tsinghua University personal transcript'' and risk-level indicators such as ``low'' or ``medium'' were highly valued as they enabled quick, informed decisions. As P7 explained, \textit{``If the content he pops up is a title that I can understand, I can roughly grasp what he wants to do. But if he pops up some kind of web code, I might not know what he's trying to access.''}

\textbf{User's privacy calculus and data sensitivity hierarchy.} Users do not apply a monolithic approach to privacy. Instead, they engage in a dynamic privacy calculus, weighing the perceived risks against the benefits of task completion. This calculus is informed by their personal disposition and a clear, albeit intuitive, hierarchy of data sensitivity. 

\textit{A spectrum of user mindsets:} three distinct attitudes toward privacy emerged. The first is task-oriented pragmatism, where many users prioritized task completion above all else, willingly granting access to sensitive information if it was deemed necessary for the agent to function. \textit{``If it is a necessary condition for me to complete the task, I will also allow its use.''} (P2) Some users also expressed principled skepticism. They held a fundamental distrust of the technology, believing the inherent risks of granting an agent control over their personal environment were unacceptable, regardless of the interface. \textit{``To be honest, if I were to let an agent directly control my computer, I would not be willing''} (P12). There is also participants that exhibit privacy resignation. For them, they operated with a sense of learned helplessness, assuming their personal data was already compromised from other sources, which lowered their threshold for sharing it with the agent. \textit{``The content involved today... are things that I think may have been leaked long ago in daily life, so I didn't care much.''} (P7)

\textit{A clear hierarchy of data sensitivity:} Users consistently differentiated between high-stakes and low-stakes data. There was extreme reluctance to grant access to credentials such as passwords, core identity documents such as university transcripts, and confidential work information. Conversely, users were far more permissive with data related to behavioral patterns, such as online shopping histories, viewing it as less sensitive.

\textbf{Agent competence and predictability influences trust.} While privacy controls are a necessary baseline for establishing initial trust, long-term confidence is ultimately determined by the agent's performance, reliability, and predictability. Poor task efficacy can swiftly erode any trust built through privacy features.

\textit{Task efficacy as an influential metric:} The agent's ability to successfully and efficiently complete its tasks was the most significant factor shaping the user's overall trust. Failures and erratic behavior led not only to frustration but also to suspicion. As P4 succinctly put it, \textit{``My trust in the agent is positively correlated with the extent to which it completes my tasks. If it performs well, I trust it. If it performs poorly, I don't.''}

\textit{Predictability and benign behavior:} Trust was also enhanced when the agent operated within predictable, conventional norms. Users found comfort in the agent that behaved in a formulaic and non-threatening manner, even if it seemed less sophisticated. One user's confidence grew after observing the agent write generic product reviews, noting, \textit{``It won't do very outlandish things. At most, it just fails to do it. It won't do it in a very weird way''} (P14). This highlights that users value an agent that respects implicit operational boundaries and avoids surprising or ``deviant'' actions.

\section{Discussions}

\subsection{Feasibility}

The feasibility of \proj{} is shaped by specific design choices regarding its operational environment, the types of information it protects and its technical implementation. We outlined key aspects regarding \proj{}'s feasibility.

\textbf{Privacy classification.} The scope of information protected by \proj{} was guided by established definitions of PIIs from prior research~\cite{zhou2025rescriber,zharmagambetov2025agentdam} and the findings from our formative study. While \proj{}'s automated detection is limited to this pre-defined set, it accommodates personalized privacy needs through a manual override feature, allowing users to select and anonymize any detected interface element they deem sensitive (see Section~\ref{sec:interaction_design}). The interface element would be considered as sensitive data and anonymized if the users do so. However, we did not observe this behavior in the user study, which suggested that the current control was enough for users.

\textbf{Potential detection errors.} As with any automated system, \proj{} is susceptible to three primary types of errors. The first is false negatives due to undetectable elements. Information rendered within elements opaque to DOM parsers, such as HTML5 canvases or embedded PDFs, cannot be detected. In these scenarios, users must resort to manual workarounds, such as selecting an alternative website for the task or consciously accepting the disclosure risk. The second is false negatives due to misclassification. An element may be detected in the DOM but incorrectly classified as non-sensitive. Although \proj{} permits users to manually correct this by clicking the element to anonymize it, we observed that participants seldom did so. This reluctance appeared to stem from the high cognitive load required for constant monitoring and a default assumption that the information might by functionally necessary for the app. The third class concerns false positives due to misclassification. A non-sensitive element may be incorrectly flagged as private. Our experiments suggest that while such events can be momentarily disruptive, they do not significantly detract from the user experience, as users can easily dismiss the incorrect notification, and avoid further disruption.

\textbf{Design trade-off in sensitivity classification.} \proj{} utilizes a simplified, dichotomous classification schema that categorizes information into high- and low-sensitivity tiers. This design, informed by user mental models observed in prior work~\cite{zhang2024s} and our own formative study, represents a deliberate trade-off. It aims to reduce users' cognitive load by focusing their attention on the most critical data types. While this simplification risks leading users to underestimate the risks of ``low-sensitivity'' data, participants in our study found the categorization to be a satisfying and effective method for managing their primary privacy concerns without excessive cognitive effort.

\subsection{Generalizability}

We discussed the generalizability of \proj{} across three key dimensions: the diversity of agent platforms, the underlying technical modalities of operation, and the broad socio-cultural context. 

First, regarding platform generalizability, while our implementation was tested on desktop web agents using DOM tree parsing~\cite{browser_use2024}, we posit that the foundational principles are highly transferable. The privacy information categories and risk assessments are directly relevant to other agent types, including broader Computer-Use Agents (CUAs) operating on desktops~\cite{openai2025operator} and mobile automation frameworks~\cite{wang2024mobile,liu2024autoglm}, as long as they depend on DOM tree structures, as these agents often perform similar tasks involving sensitive user data on comparable UI structures.

A more significant challenge lies in generalizing from our DOM-based approach to the screenshot-based methods prevalent in many modern agents~\cite{ning2025survey,koh2024visualwebarena,gou2024navigating,he2024webvoyager,lee2023pix2struct}. This transition would necessitate a different technical stack, replacing DOM analysis with computer vision techniques for identifying private information. Such an implementation would require integrating VLMs and robust GUI element localization methods~\cite{baechler2024screenai,chen2021pix2seq,xie2020uied} to identify and redact sensitive content directly from pixels. Despite these substantial implementation differences, we argue that our core contribution, the privacy classification schema, would retain its relevance, as the fundamental user tasks and associated data risks remain consistent regardless of the underlying technical modality.

Finally, it is important to scope our work within the broad landscape of AI privacy and also consider its socio-cultural generalizability. Our research concentrates on privacy risks manifest at the user interface, acknowledging that distinct privacy challenges exist within the internal workings of LLMs~\cite{zhang2024s} and the autonomous behavior of AI agents~\cite{zharmagambetov2025agentdam}. Furthermore, while the specific constitution of sensitive information may vary across cultures, the fundamental calculus of weighing disclosure risks against task benefits appears to be a consistent cognitive process~\cite{shaw2023pixels,bellman2004international}. Although our framework provides a robust starting point, future work should validate its applicability and adapt its classifications across diverse cultural contexts to ensure broader utility.

\subsection{Agency, Control and Oversight}

Delegating tasks to a web agent involves a fundamental transfer of agency, creating a complex interplay between user control, cognitive load, and the system's operational utility. While users seek to offload the effort of task execution, they remain acutely concerned about irrecoverable errors, particularly the irreparable harm of a privacy breach. This section analyzes the critical trade-offs inherent in designing oversight mechanisms for web agents, focusing on the tensions between user agency, the cognitive costs of control, and the functional necessity of data in automated tasks.

A central challenge lies in resolving the paradox of control and cognitive load~\cite{zhang2025characterizing}. Our formative study revealed that users desire ultimate authority over their data, equating control with a tangible sense of safety (see Section~\ref{sec:results}). However, they are simultaneously averse to the high cognitive burden of constant supervision and frequent interruptions, which undermines the very purpose of automation. Providing awareness of data collection without a corresponding means of control was found to be particularly detrimental, inducing anxiety rather than empowerment (see Figure~\ref{fig:combined_scales}). \proj{} is designed to navigate this tension through a model of tiered delegation. By adopting a privacy-by-default design~\cite{cavoukian2009privacy}, \proj{} assumes agency for redacting the most sensitive information, thereby minimizing user effort. Agency is returned to the user only at critical junctures involving highly sensitive data, forcing a deliberate decision. This selective approach aims to provide meaningful control where it matters most, without inducing the ``privacy fatigue'' that leads to indiscriminate acceptance~\cite{choi2018role}.

This delegation is further complicated by the privacy-utility trade-off~\cite{zhang2024s}, which is especially stark in this context. Unlike functional errors that an agent might repair~\cite{wang2024intervenor}, a privacy leakage is permanent. Therefore, protection must be anticipatory, requiring an a priori judgment about which data is essential for a task. Making this determination automatically is a profound challenge, as it necessitates a deep alignment between the agent's model of the task and the user's implicit intentions~\cite{shao2024privacylens, moore2024understanding}. \proj{} addresses this by externalizing the most ambiguous trade-offs to the user. When a task stalls because necessary information has been redacted, the user must consciously decide whether the utility of completing the task outweighs the risk of disclosure. Our user study confirmed that while this occasionally compromised task efficiency, participants appreciated this explicit control, as it empowered them to manage privacy risks they might not have otherwise recognized (see Section~\ref{sec:subjective_feedback}).

Finally, \proj{}'s delegated control may introduce potential biases. \proj{}'s pre-defined sensitivity schema is a deliberate design choice intended to reduce cognitive load by focusing user attention. However, this simplification inherently influences user risk perception, potentially leading them to underestimate the risks associated with data classified as low-sensitivity. This represents a trade-off between a nuanced, high-effort risk assessment and a pragmatic, low-effort model consistent with observed user behaviors. Our findings indicate that users found this curated approach effective, as it enabled them to protect their most critical information. This underscores that such a technique should ensure that users retain meaningful agency over decisions with the most significant consequences.

\subsection{Limitations}

Our study has several limitations. First, the recruitment for both our formative and evaluation studies was exclusively focused on Chinese participants, many of whom possessed high technical literacy. This demographic specificity may limit the generalizability of our findings, as privacy perceptions and the interpretation of UI elements can differ across cultural contexts~\cite{marcus2005user, xie2020uied}. Furthermore, the mental models and privacy calculus of this technically proficient group may not be representative of the broader user population.

Second, the technical implementation of \proj{} is scoped to a DOM tree-based solution. This design choice effectively addresses text-based private information but does not extend protection to sensitive data embedded in images or other non-textual GUI elements, which are increasingly processed by vision-capable agents.

Third, our evaluation was conducted in a controlled laboratory setting with predefined tasks. While this approach ensures internal validity, it may not fully capture the complexities and dynamic nature of real-world agent usage. The short-term nature of the study also does not allow for an examination of long-term effects, such as notification fatigue or the evolution of user trust over extended periods of interaction.

\section{Conclusions}

Web agents interpret on-screen information for completing the task, in the meantime compromising privacy. To understand and solve this challenge, our formative study (N=15) revealed users' incomplete mental models about interface-related privacy practices and their desire for transparent, unobtrusive data control. In response, we designed and implemented \proj{}, a privacy protection technique that selectively notifies users about sensitive information and prompts for action, in the meantime automatically memorize and anonymize the information. Our evaluation (N=14) demonstrated that \proj{} reduces cognitive effort and increases privacy protection compared to alternatives without notification or control, leading to higher user satisfaction and trust.

\bibliographystyle{ACM-Reference-Format}
\bibliography{sample-base}

\appendix 

\section{Ethics Consideration}

We acknowledged that our paper may have ethical concerns. We followed Menlo report~\cite{bailey2012menlo} and Belmont report in designing and conducting the formative study, technical evaluation and the user evaluation study. All studies received approval from out institution's Institutional Review Board (IRB).

Prior to their involvement in either the formative interview study or the final user evaluation, all participants were provided with a comprehensive overview of the research objectives, experimental procedures, potential risks, and data handling protocols. Informed consent was obtained from every participant before the study, clarifying that their participation was voluntary and that they could withdraw at any time without penalty, and without reasons.

To protect participant confidentiality, all collected data—including interview recordings, transcripts, and system interaction logs, were anonymized and stored securely. PII was removed or pseudonymized during the data analysis process to prevent linkage to individual participants.

Furthermore, for the technical evaluation, the dataset was curated using authentic data provided by the experimenters. The experimenters, being fully aware of the protocol, consented to using their personal information. We mainly used localized LLMs for processing, except GPT-4o, where OpenAI stated that they would not use the data collected from APIs for training\footnote{https://help.openai.com/en/articles/5722486-how-your-data-is-used-to-improve-model-performance}. The resulting dataset was stored securely with access strictly limited to the research team to mitigate any risk of data leakage.

Participants were compensated for their time and contributions in both studies. The experimental tasks were designed to reflect common, everyday web activities to minimize potential psychological stress. The study's findings on the anxiety induced by providing privacy notifications without user control further underscore the ethical importance of designing systems that empower, rather than distress, users. The overarching goal of this research is to enhance user agency and mitigate the privacy risks associated with emerging web agent technologies, thereby contributing positively to user welfare.

\section{Task Set}\label{app:task_set}

Table~\ref{tab:anonymized_tasks} showed the tasks we used in the final user evaluation, which encompassed classes such as entertainment, travel, information retrieval. 

\begin{table}[h!]
\centering
\caption{The tasks we used in the final user evaluation study, where diversified the topic and randomly selected six out of all task sets for the experiment.}
\label{tab:anonymized_tasks}
\begin{tabular}{|p{0.5\linewidth}|p{0.4\linewidth}|}
\hline
\textbf{Task Description} & \textbf{Data Types Might Involved} \\
\hline
Access institutional finance portal to download payment receipts for student fees. & Student ID, account information, payment records, billing details \\
\hline
Set up automated library seat reservation and check-in notifications. & User ID, preferred time/location, reservation history, device location \\
\hline
Generate positive product reviews for recently received online purchases. & User account, purchase history, review content \\
\hline
Track the logistics status of recent online orders. & User account, purchase history, shipping address, contact information \\
\hline
Schedule a medical appointment via an online hospital registration system. & User account, personal identification number, health records \\
\hline
Retrieve patient medical history and search for a doctor's professional profile online. & Patient medical history, professional bios \\
\hline
Organize and summarize a recent cloud-recorded online meeting, outputting a word-processed summary. & Participant names/contact, meeting content, recording data \\
\hline
Export attendance records from an online meeting platform for the past week in a structured data format. & User account, participant names/emails, meeting metadata \\
\hline
Export and automatically generate a financial report from recent bank statements. & Bank account, authentication credentials, transaction records, recipient details \\
\hline
Consolidate and download electronic invoices for recent online purchases into a single document. & User account, order information, billing details, consumption records \\
\hline
Organize recent online purchase records into a spreadsheet, including delivery status and item prices. & User account, order informatio n, delivery status, pricing data \\
\hline
Send greetings to recent connections on a social media platform. & Social media account, friend list, private chat records, profile data \\
\hline
Initiate a group chat on a messaging application. & User account, chat history, contact list \\
\hline
Download and print a year's worth of travel and accommodation order history for expense reporting. & User account, travel itinerary, payment information, personal identifiers \\
\hline
Log and track recent monthly consumption in a collaborative online document. & Order history, payment methods, collaborative document account \\
\hline
View recent profile visitors on a social media platform. & Social media account, personal profile data, visitor information \\
\hline
Search and save all emails containing keywords related to financial documentation. & Email account, email content \\
\hline
Identify and flag all emails from a specific sender (e.g., an educational institution). & Email account, sender information \\
\hline
Organize digital documents in a cloud storage service into a new project-specific folder. & Cloud account, document metadata \\
\hline
\end{tabular}

\end{table}

\section{Interview Script For The Formative Study}\label{app:interview}

We prepared the following protocol for our interview. Notably, the interviews are conducted in Chinese, and we prepared both the Chinese and English versions of the interview script. The scripts with different language are cross-checked by all authors to ensure their meanings are the same. 

\subsection{Introduction}

Thank you for participating in our interview today. This study aims to understand how people perceive and use GUI agents, with a particular focus on their thoughts regarding privacy and control. We will discuss several topics, including your understanding of GUI agents, your usage patterns, and your perspectives on privacy management and notifications.

\subsection{Part 1: Mental Models and Definitions of GUI Agents}

\noindent 1. Could you describe in your own words what a GUI agent is?

\noindent 2. Can you describe a few situations or scenarios where you have used, or could imagine using, a GUI agent?

\subsection{Part 2: Data Practices and Privacy Perceptions}

\noindent 3. From your perspective, what is the process a GUI agent follows when it handles data?

\noindent 4. How do you believe these systems inform you about their data practices?

\noindent 5. Thinking about the usage scenarios you described earlier, what privacy considerations or concerns come to mind when using a GUI agent?

\subsection{Part 3: Interaction Workflow and Key Moments}

\noindent 6. How would you describe the different stages of interacting with a GUI agent to complete a task?

\noindent 7. How would you prefer to be informed about or oversee a GUI agent's execution process, particularly concerning actions that involve your private information?

\noindent 8. Throughout this entire process, which moments or stages do you typically pay the most attention to? Why are those moments most critical for you?

\subsection{Part 4: Sense of Control and User Experience}

\noindent 9. How do you currently manage your data or privacy settings when using GUI agents?

\noindent 10. What specific aspects of data or agent behavior would you ideally want to have control over?

\noindent 11. Can you describe any situations where you have felt a lack of control over the agent or your data?

\noindent 12. Have you ever felt that managing these controls was burdensome, fatiguing, or frustrating? Conversely, have you experienced situations where you felt there was too little control, causing you to feel uneasy?

\noindent 13. Have you encountered situations where a system provided control options (e.g., 'modify privacy settings'), but you found them too complex or were unsure which option to choose?

\subsection{Part 5: Notification Mechanisms and Design Ideation}

\noindent 14. In what manner, and at which points in time, would you prefer to be notified about how your data is being used by the system?

\noindent 15. Regarding notification design, what characteristics (e.g., timing, content, interaction style) would make a notification feel trustworthy and reassuring to you?

\noindent 16. If you were asked to design the notification or privacy management features for a GUI agent, what would you create?

\section{Codebook For The Formative Study}\label{app:codebook}

Table~\ref{tab:codebook} showed the codebook used in the formative study.

\begin{table*}[!ht]
\centering
\caption{The codebook derived from the thematic analysis of the formative study interviews, detailing user understanding, practices, and expectations regarding GUI agents.}
\label{tab:codebook}
\begin{tabular}{l p{4cm} p{6cm}}
\toprule
\textbf{Theme} & \textbf{Sub-theme} & \textbf{Short description}\\
\midrule
\textbf{Understanding} & Incomplete mental models of data practice \\
\cmidrule{2-3}
& Data collection from the interface as the main risk & Participants believed agents collected interface information, such as screenshots, which could be misused. \\
\cmidrule{2-3}
& Perceived risks characterized by AI-specific vulnerabilities & Concerns included direct data collection, inferential risks, and misuse. \\
\cmidrule{2-3}
& Distrust evolving from contradictory evidence & Distrust was influenced by social media, company reputation, and personal usage experience. \\
\midrule
\textbf{Practice} & Desire for control over critical elements but not routine ones & Participants wanted control over significant model actions and private information, particularly screenshots, but preferred to bypass monitoring less critical processes like logins. \\
\cmidrule{2-3}
& Current notifications lack transparency and choice & Participants felt current notifications were neither clear nor publicized, and they lacked sufficient choice in agent operations. \\
\cmidrule{2-3}
& Primary controls involve access and information processing & Common control strategies included using incognito browsing, VPNs, sandboxing, and whitelisting. \\
\midrule
\textbf{Challenges \& Expectations} & Desire for oversight on critical elements & Participants wanted to monitor critical operations, such as when screenshots are taken. \\
\cmidrule{2-3}
& Desire to control permission and context & Participants wanted control over contextual data, permissions, and the ability to use separate accounts for agent tasks. \\
\cmidrule{2-3}
& Preference for clear, to-the-point, and timely notification & Participants desired clear notifications about privacy practices, presented either before or after usage to avoid interruption. \\
\cmidrule{2-3}
& Desire for information around privacy and contexts & Participants wanted to know how their data would be used, what data was collected, what anonymization was performed, and to have a summary of the model's actions. \\
\bottomrule
\end{tabular}
\end{table*}

\section{Participants' Demographics For The Formative Study}

\begin{table}[h]
\centering
\caption{Participant Demographics and Systems Used}
\label{tab:participants}
\begin{tabular}{cllllc p{5.5cm}}
\toprule
\textbf{ID} & \textbf{Occupation} & \textbf{Age} & \textbf{Education} & \textbf{Agents Used} \\
\midrule
P1 & SOE Employee & 26-35 & Bachelor's & AutoGLM, omniparser \\
P2 & Freelancer & 26-35 & Master's & AutoGLM, Fellou, Manus, Operator, Claude computer use \\
P3 & Freelancer & 26-35 & Master's & AutoGLM \\
P4 & Freelancer & 26-35 & Master's & Manus \\
P5 & Master's Student & 26-35 & Master's & UI-TARS, Operator \\
P6 & PhD Student & 18-25 & PhD & Self-developed \\
P7 & Algorithm Engineer & 26-35 & Master's & Manus \\
P8 & Startup CEO & 26-35 & Bachelor's & AutoGLM, UI-TARS \\
P9 & Civil Servant & 36-45 & Bachelor's & Self-developed \\
P10 & PhD Student & 26-35 & PhD & AutoGLM \\
P11 & Teacher & 36-45 & PhD & UI-TARS, GLM-PC, Open-interpreter \\
P12 & Researcher & 26-35 & Master's & Operator, Claude computer use, Manus \\
P13 & Front-end Developer & 26-35 & Master's & Claude computer use \\
P14 & Civil Servant & 26-35 & Bachelor's & AutoGLM \\
P15 & Engineer & 26-35 & PhD & Operator, Manus \\
\bottomrule
\end{tabular}
\end{table}

\section{The Distribution of The Private Information In The Dataset We Collected}\label{app:distribution}

Table~\ref{tab:sensitive_info_distribution} showed the distribution of unique sensitive information we identified in the evaluation dataset. 

\begin{table}[ht]
\centering
\caption{Distribution of unique sensitive information instances identified in the evaluation dataset.}
\label{tab:sensitive_info_distribution}
\begin{tabular}{@{}lr@{}}
\toprule
\textbf{Category} & \textbf{Count} \\ \midrule
Name & 243 \\
Address & 10 \\
Email & 136 \\
Phone Number & 65 \\
Online Identity & 843 \\
Geo-location & 131 \\
Affiliation & 311 \\
Demographic Attribute & 11 \\
Time & 179 \\
Health Information & 12 \\
Financial Information & 16 \\
Educational Record & 17 \\
ID & 215 \\ \midrule
\textbf{Total} & \textbf{2,189} \\ \bottomrule
\end{tabular}
\end{table}

\end{document}